\documentclass[twocolumn,pre,floats,superscriptaddress,longbibliography]{revtex4-1}

\usepackage[T1]{fontenc}
\usepackage[utf8]{inputenc}
\usepackage{graphicx}
\usepackage{amsmath}
\usepackage{amssymb}
\usepackage{color}
\usepackage{xspace}
\usepackage{multirow}
\usepackage{hyperref}
\usepackage{dcolumn}
\makeatletter
\newcommand*{\balancecolsandclearpage}{%
  \close@column@grid
  \cleardoublepage
  \twocolumngrid
}
\makeatother
\graphicspath{{./figures/}{./}}

\begin{document}

\title{The Quantum Transition of the Two-Dimensional Ising Spin Glass: A Tale  of Two Gaps}

\author{Massimo Bernaschi}\email{massimo.bernaschi@cnr.it}\affiliation{Istituto
  per le Applicazioni del Calcolo, CNR, Via dei Taurini 19, 00185 Rome, Italy}

\author{Isidoro González-Adalid Pemartín}\email{isidorog@ucm.es}\affiliation{Departamento de
  F\'\i{}sica Te\'orica, Universidad Complutense, 28040 Madrid, Spain}

\author{Víctor Martín-Mayor}\email{vicmarti@ucm.es}\affiliation{Departamento de
  F\'\i{}sica Te\'orica, Universidad Complutense, 28040 Madrid, Spain}

\author{Giorgio Parisi}\email{giorgio.parisi@uniroma1.it}\affiliation{Dipartimento di Fisica, Sapienza
  Università di Roma, P.le Aldo Moro 2, 00185 Rome,
  Italy}\affiliation{Nanotec-Rome unit, CNR,  P.le Aldo Moro 2, 00185 Rome, Italy}

\date{\today}

\begin{abstract}
  Quantum annealers are commercial devices aiming to solve very hard
  computational problems~\cite{johnson:11} named spin
  glasses~\cite{charbonneau:23}. Just like in metallurgic annealing one slowly
  cools a ferrous metal~\cite{kirkpatrick:83}, quantum annealers seek good
  solutions by slowly removing the transverse magnetic field at the lowest
  possible temperature. The field removal diminishes quantum fluctuations but
  forces the system to traverse the critical point that separates the
  disordered phase (at large fields) from the spin-glass phase (at small
  fields). A full understanding of this phase transition is still missing. A
  debated, crucial question regards the closing of the energy gap separating
  the ground state from the first excited state. All hopes of
  achieving an exponential speed-up, as compared to classical computers, rest on the assumption that the
  gap will close algebraically with the number of
  qspins~\cite{rieger:94b,guo:94,rieger:96b,singh:17}, but renormalization group
  calculations predict that the closing will be instead
  exponential~\cite{miyazaki:13}.  Here we solve this debate through
  extreme-scale numerical simulations, finding that both parties grasped parts
  of the truth.  While the closing of the gap at the critical point is indeed
  super-algebraic, it remains algebraic if one restricts the symmetry of
  possible excitations. Since this symmetry restriction is experimentally
  achievable (at least nominally), there is still hope for the Quantum
  Annealing paradigm~\cite{kadowaki:98,brooke:99,farhi:01}.
\end{abstract}


\maketitle

\begin{figure*}[t]%
\centering
\includegraphics[width=\textwidth]{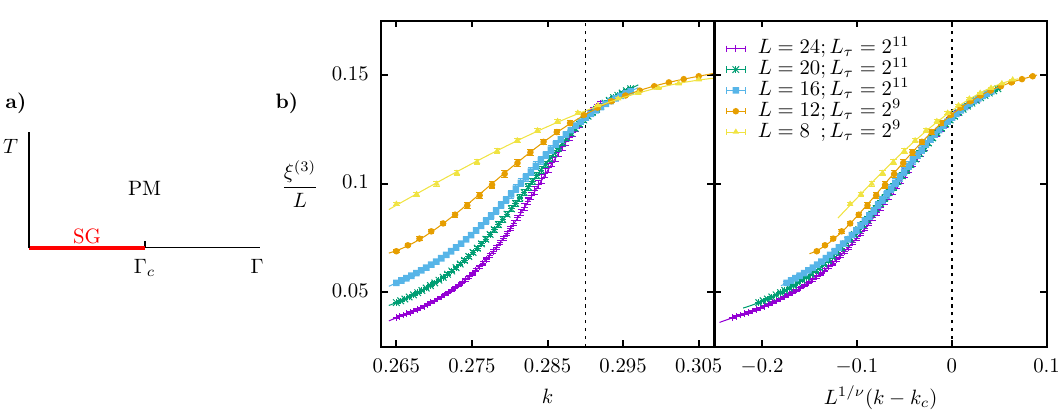}
\caption{{\bf (a)}: phase diagram for a two-dimensional Ising spin
  glass in terms of temperature $T$ and transverse field
  $\varGamma$. For all $T>0$ the system is disordered when studied at
  large lengthscales, i.e. it is in the paramagnetic phase (PM). At $T=0$,
  the Ground State seems disordered for
  $\varGamma>\varGamma_{\text{c}}$ (from the point of view of the
  computational basis). For $\varGamma<\varGamma_{\text{c}}$ we
  encounter the spin-glass phase (SG), which is different for every
  disorder realization [see Eq.~\eqref{eq:H-def}].  Panels {\bf (b)}
  illustrate our Finite-Size Scaling analysis (see,
  e.g.,~\cite{amit:05,nightingale:76}) of the critical point at
  $T=0$ and $\varGamma=\varGamma_{\text{c}}$, in terms of the parameter $k$
  that represents $\varGamma$ in the Trotter-Suzuki formulation (see
  {\bf Methods}; $k$ grows as $\varGamma$ decreases). {\bf (b)-Left:}
  correlation length $\xi^{(3)}$ in units of the lattice size $L$
  versus $k$.  The curves for the different $L$'s intersect at the
  critical point $k_{\text{c}}\approx 0.29$. {\bf (b)-Right}: Data in {\bf
    (b)-Left}, when represented as a function of the scaling variable
  $L^{1/\nu}(k-k_{\text{c}})$, with $1/\nu=0.7$, converge to a limiting curve as $L$
  grows. Errors in both {\bf (b)} panels are one standard
  deviation.}\label{fig:GS-transition}
\end{figure*}

\section{Introduction}\label{sec:Intro}

Optimization problems are ubiquitous in everyday life (think, for
instance, of deciding the best delivery route, or deciding assignments in
the national budget). These problems can be mathematically formalized:
$N$ agents (e.g. ministries seaking their part in the budget) compete,
trying to satisfy their mutually contradicting goals. The overall
frustration produced by a particular solution is quantified through a
cost function, that we attempt to minimize. This task is best solved
with the help of a computer, even for quite small $N$.  Computational
complexity studies how the computational resources (memory, computing
time, etc.) grow with $N$~\cite{papadimitriou:94}. If, for all known
algorithms, the necessary resources grow faster with $N$ than any
polynomial, e.g. like $N!$, the problem is considered \emph{hard}. A
small subset of these problems, named NP-complete, is of particular
interest: if an efficient algorithm (i.e. resources scaling
polynomially in $N$) were discovered for any of the NP-complete
problems, then a vast family of hard optimization problems would
become \emph{easy}. For physicists, the most familiar example of a
NP-complete problem is finding the minimal
energy state ---the Ground state (GS)--- of an Ising spin-glass Hamiltonian on a non-planar
graph~\cite{barahona:82b,istrail:00}.  This explains the up-surge of
hardware specifically designed for minimizing a spin-glass Hamiltonian
through a variety of algorithms and/or physical principles (see
e.g~\cite{hayato:19,matsubara:20,mcgeoch:20,mcmahon:16,janus:14}).

Specifically, the strategy that concerns us here is Quantum
Annealing. Both in the original formulation~\cite{kadowaki:98}, and also in its hardware implementation~\cite{johnson:11,mcgeoch:20}, people aim at solving spin
glasses. In particular, D-wave chips solve Ising spin glass in space dimension
$D=2$ ($D'>2$ can be coded over D-wave's $D=2$ graph~\cite{king:23};
see~\cite{baxter:08} for more on the definition of $D$).

The spin glass is the paradigmatic statistical model for
quenched disorder~\cite{parisi:94}. The Hamiltonian for $S=1/2$ spins (or qspin) is
\begin{equation}\label{eq:H-def}
H=-\frac{1}{2}\sum_{\boldsymbol{x},\boldsymbol{y}}\,\Big[
J_{\boldsymbol{x},\boldsymbol{y}}\sigma^Z_{\boldsymbol{x}}\sigma^Z_{\boldsymbol{y}}\Big]\
-\ \varGamma\sum_{\boldsymbol{x}}\,\sigma^X_{\boldsymbol{x}}\,,
\end{equation}
where the $J_{\boldsymbol{x},\boldsymbol{y}}$ are the random couplings that
define the problem instance under consideration, $\varGamma$ is the transverse
field and $\sigma^X_{\boldsymbol{x}}$ and $\sigma^Z_{\boldsymbol{x}}$ are,
respectively, the first and third Pauli matrices acting on the spin that lies
in site $\boldsymbol{x}$. The phase diagram for a two-dimensional interaction
matrix $J_{\boldsymbol{x},\boldsymbol{y}}$ is sketched in
Fig.~\ref{fig:GS-transition}--({\bf a}). For $\varGamma\!=\!\infty$, the GS has all spins as much aligned with the
transverse field as Quantum Mechanics allows them to be (paradoxically enough,
from the point of view of the computational basis that diagonalizes the
$\sigma^Z_{\boldsymbol{x}}$ matrices, this GS seems a totally random
statistical mixture). As $\varGamma$ is diminished at zero temperature, the GS
varies. In particular, at $\varGamma\!=\!0$, the GS encodes the solution of the
Optimization problem we are interested in. At some point along the annealing,
$\varGamma$ goes through the critical value $\varGamma_{\text{c}}$ that
separates the disordered GS from the spin-glass GS that does show a glassy
order in the computational basis. This is not just theorist daydreaming. In a
recent experiment conducted on a D-wave chip~\cite{king:23}, some 5000 qubits
displayed coherent quantum dynamics as $\varGamma$ went through
$\varGamma_{\text{c}}$, for annealings lasting several nanoseconds.

A strong theoretical command on the phase transition at
$\varGamma_{\text{c}}$ is clearly necessary. A very powerful tool in the analytical study of phase
transitions is the Renormalization Group (RG), which helps to clarify which
properties of the critical point are not modified by microscopic details in
which different experiments may differ. Only very broad features, such as
symmetries, matter (making it possible to classify problems into universality
classes). In fact, the study of disordered systems was one of the early applications of the RG group (see, e.g.,
Refs.~\cite{grinstein:76,parisi:79,parisi:79c}), a strategy that is firmly
established in $D=2$~\cite{cardy:96}. Yet, it has taken considerable
time and effort to show that the RG ---and the accompanying
universality--- applies to disordered systems in $D\!>\!2$ as
well~\cite{ballesteros:98b,hasenbusch:08b,fytas:13,fytas:16,fytas:19}
(even in $D\!=\!2$ this was a hard endeavor for spin
glasses~\cite{fernandez:16b}).
Unfortunately, the study of the quantum spin-glass transition at finite $D$ is a
lot behind its thermal counterpart. Essentially, only the $D\!=\!1$ case is well
understood~\cite{mccoy:68,mccoy:69,mccoy:69b,fisher:92}.

The second-simplest problem to analyze, the spin glass in
$D\!=\!2$, poses quite a challenge. Indeed, different approaches have produced
mutually contradicting predictions for the crucial physical quantity that will
ultimately decide whether or not the quantum computational complexity of the
problem to be considered is smaller than its classical counterpart.  We are
referring to the energy gap $\varDelta$ that separates the GS from the first
excited state of the Hamiltonian~\eqref{eq:H-def}. Indeed, the required
annealing time is proportional to $1/\varDelta^2$. In a spin glass with
$N\propto L^2$ qspins at $\varGamma=\varGamma_{\text{c}}$ ($L$ is the linear
size of the system), $\varDelta \propto L^{-z}$ ($z$ is the so called dynamic
critical exponent). Early Monte Carlo
simulations~\cite{rieger:94b,guo:94,rieger:96b} and a series-expansion
study~\cite{singh:17} found finite values of $z$ (e.g., $z\approx 1.5$ for
$D=2$ spin glasses~\cite{rieger:94b}). A finite $z$ is also a crucial
assumption of the droplet model for the quantum spin glass
transition~\cite{thill:95}. On the other hand, a real-space
RG analysis concludes $z=\infty$ in space dimension
$D\!=\!2$ and $3$~\cite{miyazaki:13} (a Monte Carlo simulation claims as well
$z\!=\!\infty$ in $D\!=\!2$~\cite{matoz-fernandez:16}).

The starting assumption of
Refs.~\cite{rieger:94b,guo:94,rieger:96b,king:23} was a finite value of
exponent $z$. Yet, to clarify the aforementioned controversy, our analysis
will be completely agnostic about $z$. Just as Rieger and Young pushed to
their very limit the computational capabilities at that time by using special
hardware (Transputers)~\cite{rieger:94b}, we have performed unprecedented large scale
simulations on GPUs by using highly tuned custom codes. A careful consideration of the global spin-flip symmetry,
implemented by the parity operator
$P=\prod_{\boldsymbol{x}}\sigma^X_{\boldsymbol{x}}$, turns out to be
crucial to clarify the situation.

\begin{figure*}[htb]%
\centering
\includegraphics[width=\textwidth]{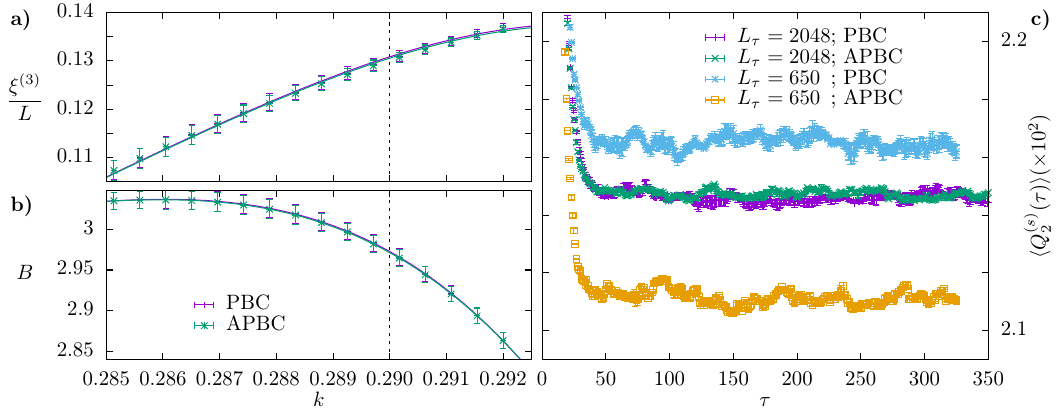}
\caption{{\bf Ensuring that the zero temperature limit has been reached
    through the comparison of periodic and antiperiodic boundary conditions
    (PBC and APBC) along the Euclidean time.} {\bf (a)}: correlation length
  $\xi^{(3)}$ (see {\bf Methods}) vs. $k$, as computed for our largest systems
  $L\!=\!24$, $L_\tau\!=\!2048$ with both PBC and APBC for the \emph{same} set
  of 1280 samples. The PBC/APBC statistical agreement indicates that the
  $T\to 0$ limit has been effectively reached for this quantity. {\bf (b)}: as
  in panel {\bf (a)} for the Binder cumulant (see {\bf Methods}). Dashed line represents the critical point, $k_c\approx 0.29$.{\bf (c)}: The even correlation
  functions $Q_2(\tau)$ defined in {\bf Methods}, as computed for a single
  sample of $L\!=\!20$ at $k=0.29$, reach rather quickly
  their large-$\tau$ plateaux, that depend both on $L_\tau$ and the boundary
  conditions. The PBC plateau \emph{decreases} upon increasing
  $L_\tau$, while the APBC significantly \emph{increases}. The reason behind
  the stronger $L_\tau$ sensitivity in the APBC case is understood (see
  {\bf  Methods}). Errors in {\bf (a)}, {\bf (b)} and {\bf (c)} are one standard
  deviate.}\label{fig:Tzero}
\end{figure*}

\section{The Ground State}

Our aim here is studying the phase transition as seen from the GS
(so that the spectra of excitations, and hence exponent $z$, does not play any
role in the analysis in Sect.~\ref{sect:Phase-transition}). This entails taking the limit $T\to 0$.

In the Trotter formulation that we use (see {\bf Methods}), the original qspins
on a $L\times L$ lattice are replaced by \emph{classical} spins on a
$L\times L\times L_\tau$ lattice, $S_{\boldsymbol{x},\tau}=\pm 1$. The extra
dimension $\tau$ is named the Euclidean time. $\varGamma$ is replaced by a new
parameter $k$, that grows as $\varGamma$ decreases.  An energy gap $\varDelta$
translates into a correlation length $\eta=1/(k\varDelta)$ along the Euclidean
time. In this formulation, the limits $T\to 0$ and $L_\tau\to\infty$ are
equivalent.

Although our main results stem from Monte Carlo simulations, a complementary
exact diagonalization effort on small systems ---see Sect.~\ref{sect:diag} and {\bf Methods}--- has being extremely useful both to shape
our analysis, and to understand how the limit $T\to 0$ is approached
(Sect.~\ref{sect:Tzero}).

\subsection{Exact diagonalization}\label{sect:diag}

The main lessons that exact diagonalization of systems with size $L\leq 6$
have taught us are the following.

The parity operator $P$ splits the spectrum of the Hamiltonian~\eqref{eq:H-def}
into even energy levels
($E_{0,\text{e}}<E_{1,\text{e}}<\ldots$) and odd levels
($E_{0,\text{o}}<E_{1,\text{o}}<\ldots$). The GS is
even and its energy is $E_{\text{GS}}=E_{0,\text{e}}$.

The first excited state is $E_{0,\text{o}}$. The minimal gap
$\varDelta=E_{0,\text{o}}-E_{0,\text{e}}$ displays dramatic fluctuations
among samples, up to the point that the statistical analysis should be conducted
in terms of $\log\varDelta$. Furthermore, $\log\varDelta$ varies significantly
with $k$. Instead, the sample-to-sample
fluctuations of the same-parity gaps,
$\varDelta_{\text{e}}\equiv E_{1,\text{e}}-E_{0,\text{e}}$ and
$\varDelta_{\text{o}}\equiv E_{1,\text{o}}-E_{0,\text{o}}$ are
\emph{very} mild (also their $k$-dependence is mild). For all our samples,
$\varDelta_{\text{e}}$ and $\varDelta_{\text{o}}$ are of similar
magnitude and, unless $\varDelta$ turns out to be inordinately large,
$\varDelta_{\text{e}},\varDelta_{\text{o}}\gg \varDelta$.

Thermal expected values of even operators (i.e., operators ${\cal A}$ such that
${\cal A}=P{\cal A}P$) reach their $T=0$ limit for a surprisingly small values of
$L_\tau$. The reasons explaining this benign behavior are understood,
see {\bf Methods}.

\subsection{The phase transition}\label{sect:Phase-transition}

The standard spin-glass correlation function, when computed on the GS (see
{\bf Methods}), is afflicted by a very large anomalous dimension that makes
the spin-glass susceptibility $\chi^{(2)}$ barely divergent at the critical
point~\cite{rieger:94b}. We have circumvented this problem by considering
instead, the correlation matrix $M$ (see {\bf Methods} and
Refs.~\cite{yang:62,sinova:00,correale:02}). From $M$, one can compute not only
$\chi^{(2)}$, but also a better behaved susceptibility $\chi^{(3)}$. The
corresponding correlation length $\xi^{(3)}$ is suitable for a standard
Finite-Size scaling study of the phase transition, see e.g.~\cite{amit:05},
that is illustrated in Figs.~\ref{fig:GS-transition}--({\bf b}).

The analysis in {\bf Methods} finds for the critical point
$k_\text{c}$, the correlation-length exponent $\nu$ and 
exponents $\gamma^{(r)}$
[$\chi^{(r)}(k_{\mathrm{c}})\sim L^{\frac{\gamma^{(r)}}{\nu}}$]:
\begin{eqnarray}
  k_{\text{c}}&=&0.2905(5)\,,\quad \frac{1}{\nu}=0.71(24)(9)\,,\label{eq:FSS1}\\
  \frac{\gamma^{(2)}}{\nu}&=& 0.27(8)(8) \,,\quad \frac{\gamma^{(3)}}{\nu}= 1.39(23)(11)\,.\label{eq:FSS2}
\end{eqnarray}
The first error estimate is statistical whereas the second error accounts for systematic effects (see {\bf Methods})
Note that the bound $\nu \geq 2/D$~\cite{chayes:86} is verified, and that
$\chi^{(2)}\sim L^{\frac{\gamma^{(2)}}{\nu}\approx 0.3}$ is, indeed, barely divergent~\cite{rieger:94b}.

\subsection{On the limit of zero temperature}\label{sect:Tzero}

The naive approach to the limit $T\to 0$ would be studying a fixed set of
samples for a sequence of growing Euclidean lengths $L_\tau$ and check when
the results become $L_{\tau}$ independent (indeed, $T>0$ effects die out as
$\text{e}^{-L_{\tau}/\eta}$). However, on the view of Sect.~\ref{sect:diag},
this is just a wishful thinking. Indeed, some instances have an inordinately
small gap (hence a huge Euclidean correlation length $\eta$), and so
$\text{e}^{-L_{\tau}/\eta}\approx 1$ for all the values of $L_\tau$ that we
can simulate (one wishes to have $\text{e}^{-L_{\tau}/\eta}\ll 1$,
instead).

Fortunately, considering simultaneously periodic and anti-periodic boundary
conditions (PBC and APBC, respectively) along the Euclidean time offers a way
out. The detailed analysis in {\bf Methods} shows that the sequence of results for
growing $L_\tau$ converges to $T\!=\!0$ \emph{from opposite sides\/}. As $L_\tau$ grows, see Fig.~\ref{fig:Tzero}-({\bf c}), the PBC
sequence monotonically decreases, while the APBC sequence increases. Thus,
statistical compatibility for both boundary conditions ensures that the
zero-temperature limit has been reached.

\begin{figure*}[htb]%
\centering
\includegraphics[width=\textwidth]{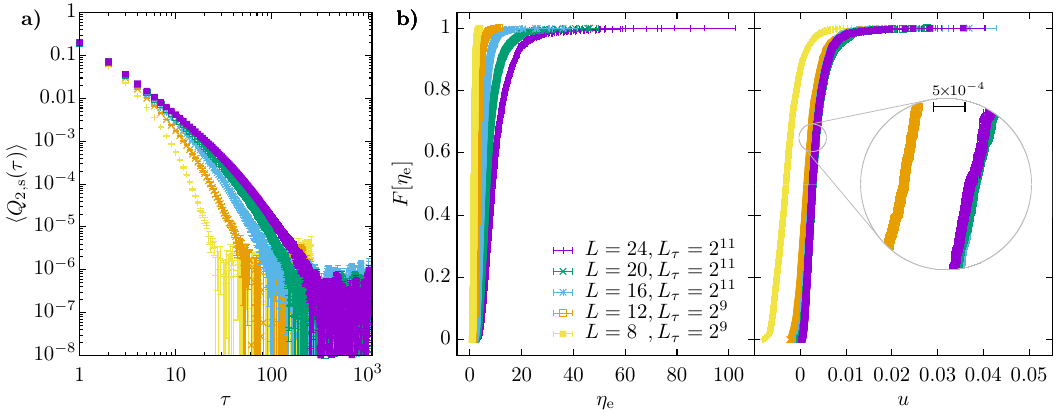}
\caption{{\bf Studying the spectra of even excitations at the critical point.}
  {\bf (a)}: The sample-averaged substracted correlation function
  $Q_{2,s}(\tau)$ (see {\bf Methods}) gets compatible with zero for moderate
  values of $\tau$, for all our system sizes. {\bf (b)-Left:} After computing the
  Euclidean correlation length $\eta_{\text{e}}^{(s)}$ for each sample, we
  compute for each $L$ the empirical distribution function
  $F(\eta_{\text{e}})$, namely the probability $F$ of finding a sample with
  $\eta_{\text{e}}^{(s)}<\eta_{\text{e}}$ (mind the horizontal error
  bars). {\bf (b)-Right:} the data in panel {\bf (b)-Left}, when plotted as a function of
  the scaling variable $u$, see Eq.~\eqref{eq:u-def}, do not show any $L$
  residual $L$ dependence but for our smallest sizes $L\!=\!8,12$.  Errors in {\bf
    (a)} and {\bf (b)} are one standard deviate.}\label{fig:even}
\end{figure*}

\section{Spectra of excitations at the transition}

The main tool to investigate excitations is the $\tau$ dependence of the
Euclidean correlation function of several operators, see {\bf Methods}.  It
will be crucial to distinguish \emph{even} operators (i.e. $P{\cal A} P={\cal A}$) from odd
operators ($P{\cal A}P=-{\cal A}$). For even operators, the decay with $\tau$ is sensitive
only to same-parity gaps (such as $\varDelta_{\text{e}}$ and
$\varDelta_{\text{o}}$ defined in Sect.~\ref{sect:diag}). Instead,
odd operators feel the different-parity energy gap $\varDelta$.

For both symmetry sectors, the correlation functions computed in a sample
decay exponentially (to zero in the case of odd operators or,
Fig.~\ref{fig:Tzero}--{\bf c}, to a plateaux for even operators). In both
cases, correlation lengths $\eta$ and the energy gaps of appropriate symmetry
are related as $\eta=1/(k\varDelta)$. Therefore, what the average over samples
of an Euclidean correlation function really features is the probability
distribution function (as computed over the different samples) of the
correlation lengths $\eta$.

From now on, we specialize to the critical point at $k_{\text{c}}\approx 0.29$.

\subsection{Even operators}

This case is of utmost relevance because only even excited states may cause
the system to leave its GS in an (ideal) quantum annealing for the
Hamiltonian~\eqref{eq:H-def}. Our approach is not entirely satisfying in this
respect because, for a given sample, we obtain the smallest of the two
same-parity gaps $\varDelta_{\text{e}}$ and
$\varDelta_{\text{o}}$ (one would like to study only
$\varDelta_{\text{e}}$). Fortunately, see {\bf Methods}, both
gaps are of similar magnitude.

The first optimistic indication comes from the (substracted) correlation
function in Fig~\ref{fig:even}--{\bf a} that indeed goes to zero (within
errors) for a moderate value of $\tau$. Indeed, the empirical distribution
function for the correlation length $\eta_\text{e}$ in
Fig~\ref{fig:even}--{\bf b} indicates mild sample-to-sample fluctuations and
a relatively weak dependence on $L$. Indeed, as shown Fig~\ref{fig:even}--{\bf
  b}, for all $L>12$, the probability
distribution function turns out to depend on the scaling variable
\begin{equation}\label{eq:u-def}
u=\frac{\eta_{\text{e}}-\eta^0_{\text{e}}}{L^{z_{\text{e}}}}\,,\ 
  \eta^0_{\text{e}}=2.2(3)\,,\ z_{\text{e}}=2.46(17)\,.
\end{equation}
(Setting $\eta^0_{\text{e}}\!=\!0$, the whole curve cannot be made to scale and the resulting estimate $z_{\text{e}}\approx 1.7$ is lower).
Thus, as anticipated, we conclude that the even symmetry sector shows
algebraic scaling for its gap.

\subsection{Odd operators}

\begin{figure*}[t]%
\centering
\includegraphics[width=\textwidth]{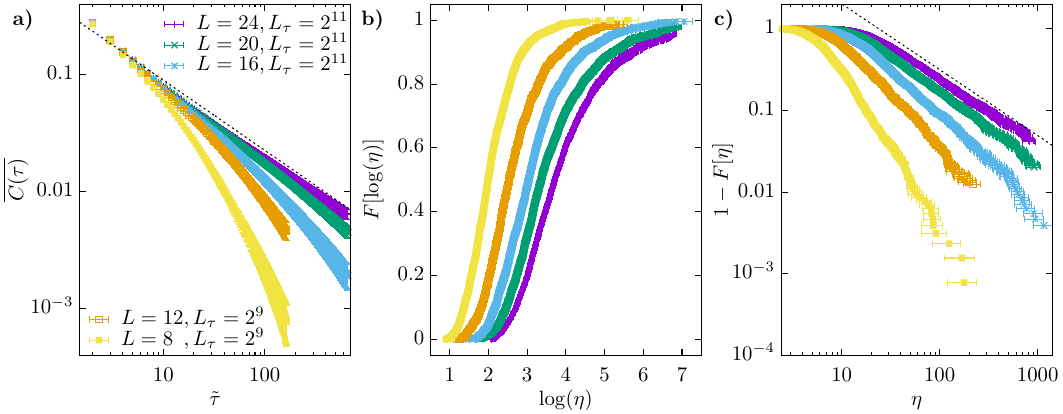}
\caption{{\bf Studying the spectra of odd operators at the critical point.}
  {\bf (a)}: The decay of the sample-averaged correlation function $C(\tau)$
  (see {\bf Methods}) approaches a power-law as $L$ increases (dashed line as a guide to the eyes). Indeed, we have
  needed to represent $C(\tau)$ in terms of
  $\tilde\tau=\frac{L_\tau}{\pi}\sin (\pi\tau/L_\tau)$ to avoid distortions
  due to the periodic boundary conditions ($\tilde\tau$ and $\tau$ are almost
  identical for small $\tau/L_\tau$). {\bf (b):} Empirical distribution
  function $F(\eta)$ as a function of $\log\eta$ for all our system
  sizes. Mind that we can compute only up to some $L$-dependent $F$ , because
  our largest $L_{\tau}$ is not large enough to allow for a safe determination
  of $\eta$ in some samples. {\bf (c):} For large $\eta$, the assymptotic
  behavior $F(\eta)=1-\frac{B}{\eta^b}$ is evinced by the linear behavior
  ---in logaritmic scale--- of $1-F$ as a function of $\eta$ (we find
  $b\!\approx\! 0.8$ ---dashed line as a guide to the eyes). Errors in {\bf
    (a)}, {\bf (b)} and {\bf (c)} are one standard deviate.}\label{fig:odd}
\end{figure*}

As it could be expected from the exact results in
$D=1$~\cite{mccoy:68,mccoy:69b,fisher:92}
and the approximate Renormalization Group for $D=2$~\cite{miyazaki:13}, the
odd correlation function $C(\tau)$ in Fig.~\ref{fig:odd}--{\bf a} displays, for
large $L$, a power law decay $C(\tau)\propto 1/\tau^{\tilde{b}}$ with
$\tilde{b}\!=\!0.6$. This implies that the magnetic susceptibility ---the
linear response to a magnetic field aligned with the Z axis--- is divergent at
the critical point. Indeed, the susceptibility diverges if $\tilde{b}<1$
(because it is twice the integral of $C(\tau)$ for $\tau$ going from $0$ to
$\infty$).

Furthermore, $\tilde{b}\!<\!1$ also for $k\!<\!k_{\text{c}}$, see {\bf Methods}. We therefore
conclude that the susceptibility is divergent in the paramagnetic phase. This
is exactly the same behaviour encountered in $D=1$. Accordingly, the
probability distribution function of the Euclidean correlation length $\eta$
---recall that $\eta=1/(k\varDelta)$--- displays a extremely fat tail, see
Figs.~\ref{fig:odd}--{\bf b} and ~\ref{fig:odd}--{\bf c}. We are in the
presence of a Levy flight, which strongly suggests that the scenario of an
infinite-randomness fixed-point~\cite{miyazaki:13} is, indeed, realized for
the $D=2$ Ising spin glass.

\section{Conclusions and Outlook}

We have solved a decades-long controversy through an extreme-scale
simulation on GPUs, and a careful consideration of the main symmetries
of the problem. Our main conclusion is very optimistic: there is no
objection of principle impeding a quantum annealer to remain in the
ground state while entering the spin-glass phase, recall
Fig.~\ref{fig:GS-transition}--({\bf a}). However, as discussed below,
this is not quite the same as solving our optimization problem.  In order
to adiabatically enter the spin-glass phase, the
annealing time would just need to grow as a power-law with the number
of qspins, recall Eq.~\eqref{eq:u-def}, provided that parity-changing
excitations are avoided (something that, at least nominally, is within
the capabilities of current hardware). Universality and the
Renormalization Group suggest that this optimistic conclusion extends
to a vast family of problems (all problems that share the
space-dimension and the basic symmetries with our spin glass on the
square lattice).

However, our findings pose as well many questions. Let us list a few.

We have seen that entering the spin-glass phase with a quantum
annealer should be doable with an effort polynomial in the number of
qbits. However, in order to solve an optimization problem, one still
needs to go adiabatically all the way from the critical point to
$\varGamma=0$. This is a difficult journey, at least for problems with
space dimension $D\to\infty$~\cite{knysh:16}. However, it has been
recently argued that an algebraic speed-up, as compared to classical
algorithms, is within reach~\cite{king:23}. Having a finite exponent
$z$ is a basic prerequisite also for algebraic speed-up.

We know that $D=2$ Ising glasses may be both hard and easy to solve on a
classical computer. For instance, problems formulated in the square lattice
with nearest-neighbor interactions can be solved quite efficiently (see,
e.g.,~\cite{khoshbakht:17}). However, adding second-neighbor interactions
results into a NP-complete problem. As far as we know, it is still unclear
whether or not these two problems belong to the same (quantum) computational
complexity-class. Since the second-neighbor interactions should play no role at
$\varGamma_{\text{c}}$, differences between the two kind of problems (if any)
should arise for transverse fields $\varGamma \!<\! \varGamma_{\text{c}}$.

In this work, we have chosen problem instances with uniform probability, but
this is not a necessity. One could focus, instead, on samples that are
particularly hard to solve with a classic digital
computer~\cite{fernandez:13,marshall:16,billoire:18}. It would be interesting
to test if the benign scaling in Eq.~\eqref{eq:u-def} remains unchanged under
these challenging circumstances. We know that these classically hard problems
are even harder to solve in a D-wave's annealer~\cite{martin-mayor:15}, but
there are many possible explanations for this poor performance of quantum
hardware (see,e.g.,~\cite{albash:17,albash:19}).

Another possible venue of research is concerned with three dimensional
systems. A recent experiment conducted on D-wave hardware suggests
$z\!\approx\!1.3$~\cite{king:23}. Whether this finite dynamic exponent refers only
to even excitations (as it would be the case in two dimensions) or it is
unrestricted~\cite{guo:94} is, probably, worthy of investigation.

\begin{acknowledgments}
We thank Antonello Scardicchio for useful discussions. We also thank Andrew King for a most useful
correspondence.
We benefited from two
EuroHPC computing grants: specifically, we had access to the Meluxina-GPU
cluster through grant EHPC-REG-2022R03-182 (158306.5 GPU computing hours) and
to the Leonardo facility (CINECA) through a LEAP grant ($2\times 10^6$ GPU
computing hours). Besides, we received a small grant (10000 GPU hours) from
the \emph{Red Española de Supercomputación}, through contract
no.~FI-2022-2-0007. Finally we thank Gianpaolo Marra for the access to the Dariah cluster in Lecce. 

This work was partially supported by Ministerio de Ciencia, Innovaci\'on y
Universidades (Spain), Agencia Estatal de Investigaci\'on (AEI, Spain,
10.13039/501100011033), and European Regional Development Fund (ERDF, A way of
making Europe) through Grant PID2022-136374NB-C21.  This research has also
been supported by the European Research Council under the European Unions
Horizon 2020 research and innovation program (Grant No. 694925—Lotglassy,
G. Parisi). IGAP was supported by the Ministerio de Ciencia, Innovaci\'on y
Universidades (MCIU, Spain) through FPU grant No.~FPU18/02665.
  
\end{acknowledgments}

\section*{Conflict of interests}
The authors declare no conflict of interests.

\section*{Authors' contributions}
G.P. suggested to undertake this project.
M.B., I.G.-A.P. and V.M.-M. planned research.
M.B., I.G.-A.P., V.M.-M., and G.P. contributed computer code.
M.B. and I.G.-A.P. carried out the simulations.
M.B., I.G.-A.P., V.M.-M., and G.P. analyzed the data and wrote the paper.

\section*{Availability of data} Data can be obtained from the corresponding
author (I.G.-A.P.) upon reasonable request.

\section*{Code availability} The authors will make their code publicly
available through a separate publication. Meanwhile, upon reasonable
request, the code can be obtained from M.B.

\appendix
\section{Methods}

\subsection{Model and simulations}
Our qspins occupy the nodes of a square lattice of side $L$, endowed
with periodic boundary conditions. The coupling matrix
$J_{\boldsymbol{x},\boldsymbol{y}}$ in Eq.~\eqref{eq:H-def} is non-vanishing only for nearest lattice-neighbors. 
A problem instance, or
sample, is defined by the choice of the
$J_{\boldsymbol{x},\boldsymbol{y}}$ matrix. The non-vanishing matrix
elements,
$J_{\boldsymbol{x},\boldsymbol{y}}=J_{\boldsymbol{y},\boldsymbol{x}}$
are random, independent variables in our simulations. Specifically, we
chose $J_{\boldsymbol{x},\boldsymbol{y}}=\pm J_0$ with $50\%$
probability. We chose energy units such that $J_0=1$.

Given an observable ${\cal A}$, we shall refer to its thermal
expectation value in a given sample as $\langle{\cal A}\rangle$, see
below
Eq.~\eqref{eqM:Trotter-transfer-matrix} (the
temperature is as low as possible, ideally $T\!=\!0$). Thermal expectation values are averaged over the choice of couplings
(\emph{quenched} disorder, see e.g.~\cite{parisi:94}). We shall denote
the second average ---over disorder--- as
$\overline{\langle{\cal A}\rangle}$.

\subsubsection{Crucial symmetries}\label{sect:symmetries}

\begin{figure}[htb]
\includegraphics[width=\columnwidth]{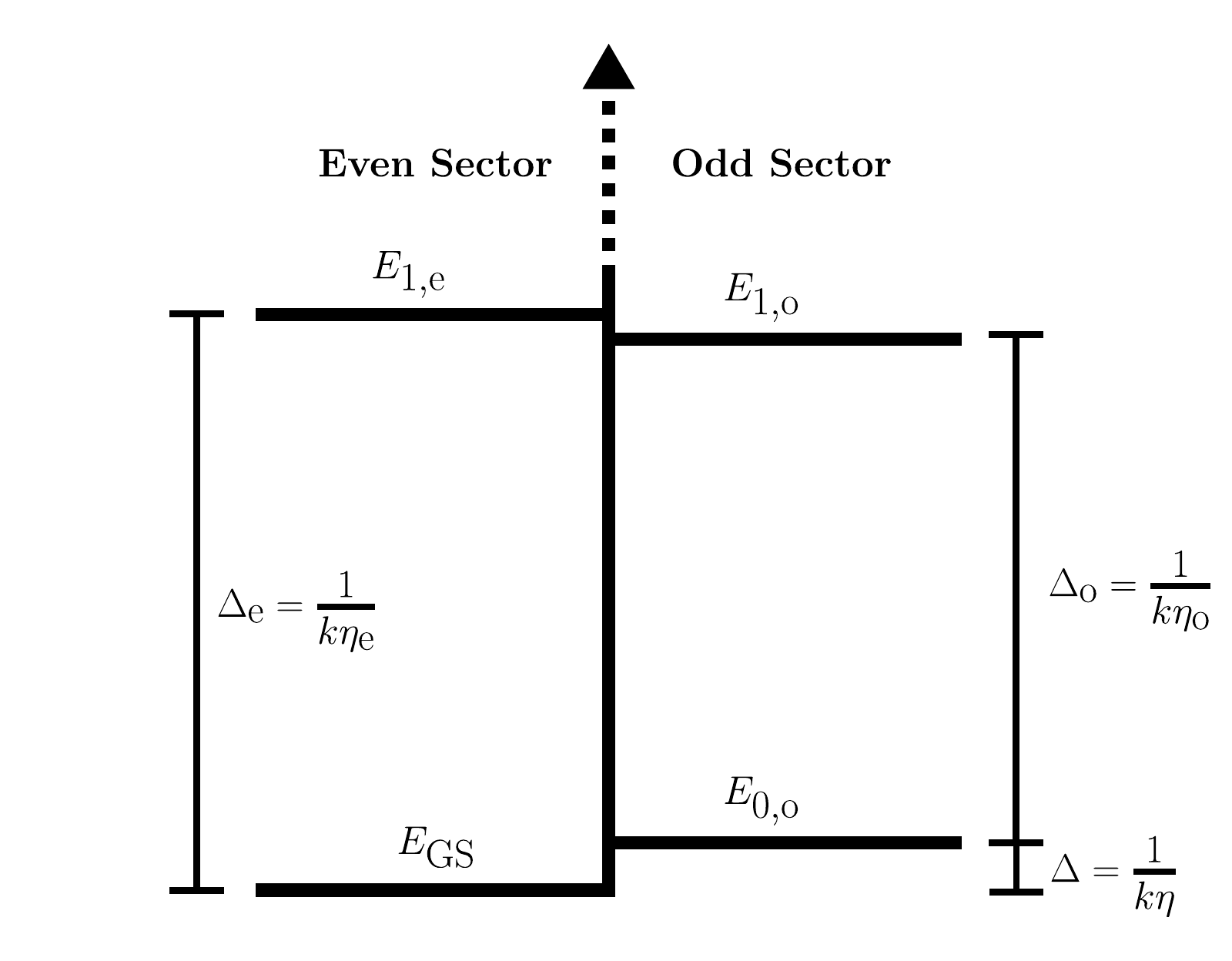}
\caption{Schematic representation of the energy spectrum. As it is explained in the main text, the parity symmetry splits the spectra into even and odd sectors according to the parity of states. We shall name the even eigenvectors of the transfer matrix~\eqref{eqM:Trotter-transfer-matrix} as $|0_{\text{e}}\rangle$, $|1_{\text{e}}\rangle$, \ldots, with corresponding eigenvalues $\text{e}^{-kE_{\text{GS}}}$, and $\text{e}^{-k(E_{\text{GS}}+\Delta_{n,\text{e}})}$ for $n=1,2,3,\ldots$ [we use the shorthand $\Delta_{\text{e}}=\Delta_{1,\text{e}}$].  For the odd sector, we have $|0_{\text{o}}\rangle$, $|1_{\text{o}}\rangle$, \ldots with eigenvalues $\text{e}^{-k(E_{\text{GS}}+\Delta)}$, and $\text{e}^{-k(E_{\text{GS}}+\Delta+\Delta_{n,\text{o}})}$ for $n=1,2,3,\ldots$ [we use the shorthands $\Delta=E_{0,\text{o}}-E_{\text{GS}}$, and $\Delta_{\text{o}}=\Delta_{1,\text{o}}$]. Notice that expectation values at $T\!=\!0$ are determined solely by $|0_{\text{e}}\rangle$.}
\label{fig_M:spectrum}
\end{figure}

The most prominent symmetries in this problem are the gauge and the
parity symmetries. Both symmetries are exact for the
Hamiltonian~\eqref{eq:H-def} and for its Trotter-Suzuki approximation
(see Sect.~\ref{sectM:Trotter-Suzuki}).

The parity symmetry
$P=\prod_{\boldsymbol{x}}\sigma^X_{\boldsymbol{x}}$ is a
self-adjoint, unitary operator that commutes with the
Hamiltonian~\eqref{eq:H-def}, as well as with the
exact~\eqref{eqM:true-transfer} and
approximate~\eqref{eqM:Trotter-transfer-matrix} transfer matrices.  The
Hilbert space is divided into two sub-spaces of the same dimension,
according to the parity eigenvalue, either $+1$ (even states) or $-1$
(odd states). We classify also operators as either \emph{even}
(i.e. $P{\cal A} P={\cal A}$) or odd operators
($P{\cal A}P=-{\cal A}$). Matrix elements of even operators can be
non-vanishing only if the two states have the same parity (on the contrary, for odd
operators the parity of the states should differ). An
oversimplified but enlightening cartoon of the spectra in our problem
is represented in Fig.~\ref{fig_M:spectrum} (see below some
exact-diagonalization results that support this view).

The parity symmetry is just a particular case of gauge
transformation.  Let us arbitrarily choose for each site 
$n_{\boldsymbol{x}}=0$ or 1. The corresponding gauge operator
$G_{\{n_{\boldsymbol{x}}\}}=\prod_{\boldsymbol{x}}(\sigma^X_{\boldsymbol{x}})^{n_{\boldsymbol{x}}}$
is self-adjoint and unitary. It transforms the Hamiltonian in
Eq.~\eqref{eq:H-def} in a Hamiltonian of the same type, but with
modified couplings~\cite{toulouse:77}:
$J_{\boldsymbol{x},\boldsymbol{y}}\longrightarrow
J_{\boldsymbol{x},\boldsymbol{y}}
(-1)^{n_{\boldsymbol{x}}+n_{\boldsymbol{y}}}\,.$ The gauge symmetry is enforced by the process of
taking the disorder average. Indeed, the gauge-transformed coupling
matrix has the same probability as the original one. Hence, meaningful
observables should be invariant under an arbitrary gauge
transformation. The parity operator is obtained by setting
$n_{\boldsymbol{x}}=1$ for all sites, which does not modify the
$J_{\boldsymbol{x},\boldsymbol{y}}$ (hence, parity is a symmetry for a
given problem instance not just a symmetry induced by the disorder
average).

\subsubsection{The Trotter-Suzuki formula}\label{sectM:Trotter-Suzuki}

We follow the Trotter-Suzuki approximation~\cite{trotter:59,suzuki:76}
that replaces the original qspins on a $L\times L$ lattice by
\emph{classical} spins on a $L\times L\times L_\tau$ lattice,
$S_{\boldsymbol{x},\tau}=\pm 1$. The extra dimension $\tau$ is named
 \textit{Euclidean time}. We shall write $\boldsymbol{S}$ as a shorthand for the $L^DL_\tau$ spins in the system ($D\!=\!2$, here). Instead,
$\boldsymbol{S}_\tau$ will refer to the $L^D$ spins at time $\tau$.
The probability of $\boldsymbol{S}$ is given by
\begin{equation}\label{eqM:P-S-tau}
p(\boldsymbol{S}) = \frac{\text{e}^{-k\mathcal{E}(\boldsymbol{S})}}{Z}\,,\quad Z=\sum_{\lbrace\boldsymbol{S}\rbrace} \text{e}^{-k\mathcal{E}(\boldsymbol{S})}\,,
\end{equation}
with [$Z$ in Eq.~\eqref{eqM:P-S-tau} is named the partition function]
\begin{eqnarray}\label{eqM:Trotter-action}
\mathcal{E}(\boldsymbol{S})&=& -\sum_{\tau=0}^{L_{\tau-1}}\Big[\frac{1}{2}\sum_{\boldsymbol{x},\boldsymbol{y}}\,
J_{\boldsymbol{x},\boldsymbol{y}} S_{\boldsymbol{x},\tau}S_{\boldsymbol{y},\tau}\\\nonumber &+& \sum_{\boldsymbol{x}} S_{\boldsymbol{x},\tau} S_{\boldsymbol{x},\tau+1}\Big]\,,\ \varGamma=\frac{-1}{2k}\log\text{tanh} k\,.
\end{eqnarray}
Periodic boundary conditions (PBC) are assumed along the
Euclidean time. Below, we shall find it useful to consider as well
antiperiodic boundary conditions \emph{only} along the $\tau$ direction (APBC). Besides,
as the reader may check, $k$ is a monotonically decreasing function of
$\varGamma$.

Maybe the most direct connection between the $D+1$ classical spin
system and the original quantum problem is provided by the transfer
matrix $\tilde{\cal T}$~\cite{kogut:79,parisi:88}.  Let us define
$\mathcal{H}_0\!=\!-\frac{1}{2}\sum_{\boldsymbol{x},\boldsymbol{y}}J_{\boldsymbol{x},\boldsymbol{y}}\sigma^Z_{\boldsymbol{x}}\sigma^Z_{\boldsymbol{y}}$
and
$\mathcal{H}_1\!=\!-\varGamma\sum_{\boldsymbol{x}}\,\sigma^X_{\boldsymbol{x}}$. The
quantum thermal expectation value at temperature $T\!=\!1/(k L_\tau)$ is
\begin{equation}\label{eqM:true-transfer}
  \langle\langle {\cal A}\rangle\rangle =\frac{\text{Tr}\,{\cal A}\,\tilde{\cal T}^{L_\tau}}{\text{Tr}\, \tilde{\cal T}^{L_\tau} }\,,\  \tilde{\cal T} = \text{e}^{-k(\mathcal{H}_0+\mathcal{H}_1)}\,.
\end{equation}
Now, for  ${\cal A}\!=\!A_{\text{cl}}(\{\sigma^Z_{\boldsymbol{x}}\}), A_{\text{cl}}$ being an arbitrary function, the Trotter-Suzuki approximation amounts
to substitute the true transfer matrix in
Eq.~\eqref{eqM:true-transfer} by its proxy ${\cal T}$ [${\cal T}=\tilde{\cal T} + {\cal O}(k^3)$]:
\begin{equation}\label{eqM:Trotter-transfer-matrix}
\langle {\cal A}\rangle =\frac{\text{Tr}\,{\cal A}\,{\cal T}^{L_\tau}}{\text{Tr}\, {\cal T}^{L_\tau} }\,,\ 
{\cal T} = \text{e}^{-\frac{k}{2}\mathcal{H}_0}\text{e}^{-k\mathcal{H}_1}\text{e}^{-\frac{k}{2}\mathcal{H}_0}\,.
\end{equation}
$\langle {\cal A}\rangle$ can be computed as well by averaging 
$A_{\text{cl}}(\boldsymbol{S}_\tau)$,
evaluated over configurations distributed according to \eqref{eqM:P-S-tau} (the value of $\tau$ is arbitrary, hence one may gain statistics by averaging over $\tau$).

Finally, let us emphasize that both ${\cal T}$ and
$\tilde{\cal T}$ are self-adjoint, positive-define  transfer matrices that share
the crucial symmetries discussed in Sect.~\ref{sect:symmetries}.

\subsection{Observables}

The quantities defined in Sect.~\ref{sectM:one-time} aim at probing the ground state as $k$ [and hence $\varGamma$~\eqref{eqM:Trotter-action}] varies. These quantities will always be averaged over disorder before we proceed with the analysis. 

Instead, the time correlations in Sect.~\ref{sectM:two-times} will probe the excitations. These time correlations will be analyzed individually for each sample (sample-to-sample fluctuations are considered in Sect.~\ref{sectM:fits-and-distributions}).

\subsubsection{One-time observables}\label{sectM:one-time}
We consider the $L^D\times L^D$ correlation matrices $M$ and $\hat{M}$~\cite{yang:62,sinova:00} ---$\boldsymbol{p}\!=\!(2\pi/L,0)$ or $(0,2\pi/L)$:
\begin{equation}\label{eqM:def-M}
M_{\boldsymbol{x},\boldsymbol{y}} =\langle\sigma^Z_{\boldsymbol{x}}\sigma^Z_{\boldsymbol{y}}\rangle\,,\quad [\hat M]_{\boldsymbol{x},\boldsymbol{y}} = M_{\boldsymbol{x},\boldsymbol{y}} \text{e}^{\text{i} \boldsymbol{p} \cdot (\boldsymbol{x} -\boldsymbol{y})}\,.
\end{equation}
The $r$-body spin-glass susceptibilities at both zero and minimal momentum are 
\begin{equation}\label{eqM:chi-r}
\chi^{(r)}= \frac{\overline{\text{Tr} \big[M^r\big]}}{L^D}\,,\ F^{(r)}= \frac{1}{L^D}\overline{ \text{Tr} \big[\hat{M}M^{r-1}\big]}\,.
\end{equation}
$\chi^{(r)}$ and $F^{(r)}$ give us access to the second-moment correlation length (see, e.g., Ref.~\cite{amit:05})
\begin{equation}\label{eqM:xi-r}
\xi^{(r)}=\frac{1}{2\sin(\pi/L)}\sqrt{\frac{\chi^{(r)}}{F^{(r)}}-1}\,.
\end{equation}
As $L$ grows, $\chi^{(r)}$ and $\xi^{(r)}$ remain of order 1 in the paramagnetic phase whereas, in the critical region, they diverge as $\chi^{(r)}\sim L^{\gamma^{(r)}/\nu}$  and $\xi^{(r)}\sim L$.
In the spin-glass phase, $\chi^{(r)}\sim L^{D(r-1)}$  ($\xi^{(r)}\sim L^a$ with some unknown exponent $a>1$). 
 
Our $\chi^{(r=2)}$ and $\xi^{(r=2)}$ are just the standard quantities in the spin-glass literature~\cite{palassini:99,ballesteros:00}. In fact, in the simplest approximation ---see Ref.~\cite{correale:02} for a more paused exposition--- at criticality and  for large separations $r$ between $\boldsymbol{x}$ and $\boldsymbol{y}$, $M_{\boldsymbol{x},\boldsymbol{y}}\sim v_{\boldsymbol{x}}v_{\boldsymbol{y}}/r^a$ with $v_{\boldsymbol{x}}, v_{\boldsymbol{y}}\sim 1$ (so, $\gamma^{(r)}/\nu=(r-1)D -r a$ in this approximation). Hence, if $D>a$, $\gamma^{(r)}$ grows with $r$. Indeed, $r\!=\!3$ turns out to be a good compromise between statistical errors, that grow with $r$, and a strong enough critical divergence of $\chi^{(r)}$ 
($\chi^{(r=2)}$ barely diverges~\cite{rieger:94b}).

Besides, we have computed the Binder cumulant as ($Q_2=L^D\chi^{(r=2)}$)
\begin{equation}\label{eqM:Binder}
B=\frac{Q_4}{Q_2^2}\,,\ Q_4=\sum_{\boldsymbol{x},\boldsymbol{y},\boldsymbol{z},\boldsymbol{u}}\overline{\langle \sigma^Z_{\boldsymbol{x}}\sigma^Z_{\boldsymbol{y}} \sigma^Z_{\boldsymbol{z}}\sigma^Z_{\boldsymbol{u}}\rangle^2}\,. 
\end{equation}
The Gaussian nature of the fluctuations in the paramagnetic phase causes $B$ to approach 3 as $L$ grows for fixed $k\!<\!k_\text{c}$. $B$ reaches different large-$L$ limits for fixed $k\!\geq\!k_\text{c}$ 
(for $k\!>\!k_\text{c}$ different behaviors are possible, depending on the degree of Replica Symmetry Breaking~\cite{mezard:87}).

\subsubsection{Two-times observables}\label{sectM:two-times}
Let us start by defining the time-correlation function of an observable ${\cal A}$ (for simplicity, consider a product of $\sigma^Z$ operators at some sites):
\begin{equation}\label{eqM:C_A_transfer}
C_{\cal A}(\tau)=\frac{\text{Tr}\,{\cal A}\,{\cal T}^{\tau}\,{\cal A}\,{\cal T}^{L_\tau-\tau}}{\text{Tr}\, {\cal T}^{L_\tau} }\,.
\end{equation}
$C_{\cal A}(\tau)$  can be computed 
from our spin configurations distributed according to the classical
weight \eqref{eqM:P-S-tau} by averaging $\sum_{\tau_1=0}^{L_\tau-1}
A_{\text{cl}}(\boldsymbol{S}_{\tau_1}) A_{\text{cl}}(\boldsymbol{S}_{\tau_1+\tau})/L_\tau^2$ (notations as in Sect.~\ref{sectM:Trotter-Suzuki}).

Specifically,  we have considered
\begin{equation}\label{eqM:def-C-Q2-tau}
C(\tau)=\frac {\sum_{\boldsymbol{x}} C_{\sigma^Z_{\boldsymbol{x}}}(\tau)}{L_\tau^D}\,,\ Q_2(\tau)= \frac{\sum_{\boldsymbol{x},\boldsymbol{y}} C_{\sigma^Z_{\boldsymbol{x}}\sigma^Z_{\boldsymbol{y}}}(\tau)}{L_\tau^{2D}}\,.
\end{equation}

\begin{figure}[htb]
    \centering
    \includegraphics[width=\columnwidth]{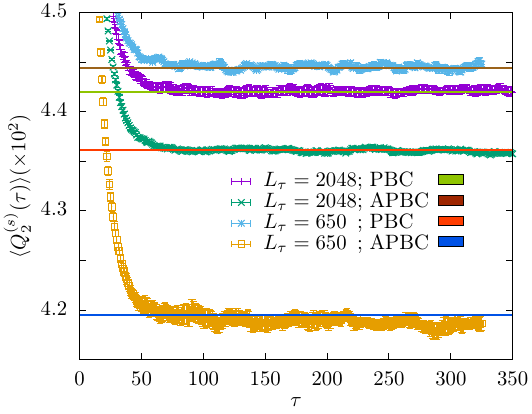}
    \caption{Even correlation functions $Q_2(\tau)$ defined in Eq.~\eqref{eqM:def-C-Q2-tau}, as computed for a single sample of $L=20$ at $k=0.29\approx k_{\text{c}}$. The corresponding $Q_2$ value calculated from $\text{Tr} M^2/L^{2D}$ is represented by a complementary colored horizontal line.}
    \label{fig_M:Q2_tau}
\end{figure}

Let us briefly recall some general results~\cite{kogut:79,parisi:88}
about $C_{\cal A}(\tau)$, that follow from the spectral decomposition of the transfer matrix (to simplify notations, let us first disregard the parity symmetry and consider PBC). 

The $\tau$-dependence is given by the additive contribution of every pair of states $E_n<E_m$ ($n=0$ is the Ground State). Every pair generates an exponentially decaying term
$B_{n,m}[\text{e}^{-\tau/\eta_{n,m}}+\text{e}^{-(L_\tau-\tau)/\eta_{n,m}}]$ with correlation length $\eta_{n,m}=1/(k \Delta_{n,m})$, $\Delta_{n,m}=(E_m-E_n)$. The amplitude is
$B_{n,m}=\text{e}^{-L_\tau/\eta_{0,n}}|\langle n| A|m\rangle|^2/\hat Z$, with $\hat Z=1+\sum_{n>0} \text{e}^{-L_\tau/\eta_{0,n}}\,$. Hence
if $L_\tau\gg \eta_{0,n}$ the contribution of this pair of states can be neglected.  Besides, in the presence of parity symmetry, for an even ${\cal A}$ we find $B_{n,m}=0$ if the parity of $|n\rangle$ and $|m \rangle$ differ (for odd operators $B_{n,m}=0$ if both parities are equal). This is why the largest correlation length for $Q_2(\tau)$ is the maximum of $\eta_{\text{e}}$ and $\eta_{\text{o}}$, whereas the relevant correlation length for $C(\tau)$ is $\eta$ (recall Fig.~\ref{fig_M:spectrum}). 

Moreover, for even operators, every state $|n\rangle$ provides an additive contribution to a $\tau$-independent term (namely, the plateau in Fig.~\ref{fig_M:Q2_tau}): 
$\text{e}^{-L_\tau/\eta_{0,n}}|\langle n| A|n\rangle|^2/\hat Z$. Instead, for odd
operators $|\langle n| A|n\rangle|\!=\!0$ (hence, odd operators lack a plateau).
To manage the case of APBC, one just needs to add an extra parity operator as a final factor in both the numerator and the denominator of both Eqs.~(\ref{eqM:Trotter-transfer-matrix}, \ref{eqM:C_A_transfer}). If parity is a symmetry, as is the case for our problem,
$\hat Z$ is modified as $\hat Z=1+\sum_{n>0} p_n \text{e}^{-L_\tau/\eta_{0,n}}\,$ ($p_n=\pm 1$ is the parity of the state) and the contribution to the APBC plateau gets an extra factor $p_n$ , as well. 

\subsubsection{The limit of zero temperature}\label{sectM:zero-temperature}

We shall assume that we can reach $L_\tau$ large enough to have
$\text{e}^{-L_\tau/\eta_{\text{e}}},\text{e}^{-L_\tau/\eta_{\text{o}}}\ll 1$ (notations are explained in Fig.~\ref{fig_M:spectrum}). Instead, we shall \emph{not} assume that $\epsilon\equiv\text{e}^{-L_\tau/\eta}$ is small (in fact, for some samples one could even have  $\epsilon\approx 1$). 

Now, consider an even operator ${\cal A}$, and let us define ${\cal A}_{\text{e}}=\langle 0_{\text{e}}|{\cal A}| 0_{\text{e}}\rangle$
and ${\cal A}_{\text{o}}=\langle 0_{\text{o}}|{\cal A}| 0_{\text{o}}\rangle$ (the thermal expectation value at exactly $T\!=\!0$ is ${\cal A}_{\text{e}}$). The plateau at  $\tau\gg \eta_{\text{e}},\eta_{\text{o}}$, see Fig.~\ref{fig_M:Q2_tau}, is given ($\zeta=1$ for PBC and $\zeta=-1$ for APBC)
\begin{equation}\label{eqM:plateau}
C_{\cal A}(\tau\gg \eta_{\text{e}},\eta_{\text{o}})= {\cal A}_{\text{e}}^2\ +\ [{\cal A}_{\text{o}}^2 - {\cal A}_{\text{e}}^2]\frac{\zeta\epsilon}{1+\zeta\epsilon}\,.
\end{equation}
Thus, we get for  the plateau of $Q_2(\tau)$
\begin{equation}\label{eqM:plateau-2}
Q_2(\tau\gg \eta_{\text{e}},\eta_{\text{o}})= Q_{2,\text{e}}+[Q_{2,\text{o}} - Q_{2,\text{e}}]\frac{\zeta\epsilon}{1+\zeta\epsilon}\,,
\end{equation}
where  $Q_{2,\text{e}}$ and $Q_{2,\text{o}}$ are, respectively, the average over all pairs  $(\boldsymbol{x},\boldsymbol{y})$ of ${\cal A}_{\text{e}}$ and ${\cal A}_{\text{o}}$ [${\cal A}=\sigma^Z_{\boldsymbol{x}} \sigma^Z_{\boldsymbol{y}}$, recall Eq.~\eqref{eqM:def-C-Q2-tau}]. To an excellent numerical accuracy, the l.h.s. of Eq.~\eqref{eqM:plateau-2} is also the value one gets for $\text{Tr} M^2/L^{2D}$, see Fig.~\ref{fig_M:Q2_tau}. As a matter of fact, the difference between $\langle{\cal A}\rangle^2$ and its plateau is $\zeta\epsilon({\cal A}_{\text{e}}-{\cal A}_{\text{o}})^2/(1+\zeta\epsilon)^2$ [hence, quadratic in $({\cal A}_{\text{e}}-{\cal A}_{\text{o}})$ rather than linear as in Eq.~\eqref{eqM:plateau}].

Now, in spite of their simplicity, two important consequences follow from Eqs.~(\ref{eqM:plateau}, \ref{eqM:plateau-2}). 

First, the limit $T\to 0$ (or $L_\tau\to\infty$) is approached monotonically. Furthermore,  the systems with PBC and  APBC [Fig.~\ref{fig_M:Q2_tau}] approach the limit from opposite sides. We have explicitly checked all our samples, finding no instance where the APBC plateau lies above the PBC one (it is intuitively natural to expect that the PBC system will be more ordered than the APBC one). Hence, we conclude that the samples with PBC converge to $T\to 0$ from above, whereas the APBC ones converge from below. 

Second,
since $Q_2(\tau)$ is bounded between 0 and 1 also for APBC, we conclude that  $|Q_{2,\text{o}} - Q_{2,\text{e}}|< (1-\epsilon)/\epsilon$. Hence, quite paradoxically, the particularly difficult samples with $\epsilon\approx 1$ generate a very small finite-temperature bias in the \emph{\textit{PBC}} estimator [compare the $L_\tau$ dependence of the PBC and the APBC plateaux in Fig.~\ref{fig_M:Q2_tau}]. This is why we are able to reach the $T\to 0$ limit for the even operators, even if a fraction of our samples suffer from a large value of $\epsilon$.

\subsubsection{Simulation details}

We have followed two approaches: exact diagonalization of the transfer matrix~\eqref{eqM:Trotter-transfer-matrix} and Markov Chain Monte Carlo simulations of the classical weight~\eqref{eqM:P-S-tau}. GPUs were crucial for both. We provide here only the main details (the interested reader is referred to~\cite{bernaschi:23}).

Exact diagonalization is limited to small systems (up to  $L\!=\!6$ in our case). Indeed, the
transfer matrix has a size $2^{L^2}\times 2^{L^2}$. Parity symmetry has allowed us to
represent ${\cal T}$ as a direct sum of two sub-matrices of half that size~\cite{bernaschi:23}. Specifically, we have computed the eigenvalues
$\text{e}^{kE_{0,\text{e}}}$, $\text{e}^{kE_{1,\text{e}}}$, $\text{e}^{kE_{0,\text{o}}}$, $\text{e}^{kE_{1,\text{o}}}$, as well as the corresponding eigenvectors $|0_\text{e}\rangle$,
$|0_\text{o}\rangle$, $|0_\text{e}\rangle$,
$|0_\text{o}\rangle$, for 1280 samples of $L\!=\!6$ at $k\!=\!0.31$ and $0.305$ (the same
samples at both $k$ values). We repeated the calculations for a subset of 350 samples at $k\!=\!0.3$ and $0.295$. 
We managed to keep the computing time within an acceptable time frame of 20 minutes \textit{per} diagonalization using 256 GPU, thanks to a highly tuned custom matrix-vector product~\cite{bernaschi:23}.
These computations have proven to be invaluable in the process of taking the limit  $L_\tau\to\infty$. (see Sect.~\ref{sectM:exact-diagonalization})

Our Monte Carlo simulations employed the Parallel Tempering algorithm~\cite{hukushima:96}, implemented over the $k$ parameter in~\eqref{eqM:P-S-tau}, to  ensure equilibration. We
equilibrated 1280 samples of every of our lattice sizes (see Table~\ref{tabM:sim_info}).
As a rule, we have estimated errors using the bootstrap method~\cite{efron:94}, as applied to the disordered average.

We have simulated 6 real replicas of every sample (i.e. six statistically independent simulations of the system), for multiple reasons.  Replicas allow us to implement the equilibration tests based on the tempering dynamics~\cite{billoire:18}. They also provide unbiased estimators of products of thermal averages~\eqref{eqM:chi-r}. Finally, fluctuations between replicas allow us to estimate errors for the time correlation functions~\eqref{eqM:def-C-Q2-tau}, as computed in a single sample (see Sect.~\ref{sectM:fits-and-distributions}).

The Monte Carlo code exploits a three-level parallelization (multispin coding, domain decomposition, parallel tampering) that allows keeping the spin-update time below 0.5 picoseconds~\cite{bernaschi:23}, competitive with dedicated hardware~\cite{janus:14}. 

\begin{table}[htb]
    \renewcommand{\arraystretch}{1.5}
\begin{tabular}{|c|c|c|c|c|c|}
\hline
$L$ & $L_\tau$ & $k_{\text{min}}$ & $k_{\text{max}}$ & $N^{\circ}$ of $k$ & MC steps\\
\hline
$8$ & $2^{9}$ & $0.265$ & $0.305$ & $16$ & $4.5\!\times\! 10^7$\\
$12$ & $2^{9}$ & $0.265$ & $0.305$ & $24$ & $10.5\!\times\! 10^7$ \\
$16$ & $2^{11}$ & $0.265$ & $0.295986$ & $48$ & $50.1\!\times\! 10^7$\\
$20$ & $2^{11}$ & $0.265$ & $0.295986$ & $56$ & $67.8\!\times\! 10^7$\\
$24$ & $2^{11}$ & $0.265$ & $0.292$ & $60$ & $78\!\times\! 10^7$\\
\hline
\end{tabular}
\caption{Simulation parameters for the different system sizes. The $k$ ranges have been chosen to ensure the critical point $k_\text{c}$ belonged to the range [see Eq~\eqref{eq:FSS1}]. The $N^{\circ}$ values of $k$, $k_{\text{min}}\leq k\leq k_{\text{max}}$, are uniformly distributed. We also provide the number of Metropolis sweeps performed (an elementary step consisted of 30 full-lattice Metropolis sweeps, followed by a Parallel Tempering attempt of exchanging the $k$ value.}
\label{tabM:sim_info}
\end{table}

\subsection{Exact diagonalization}\label{sectM:exact-diagonalization}

\begin{figure*}[t]
\includegraphics[width=\textwidth]{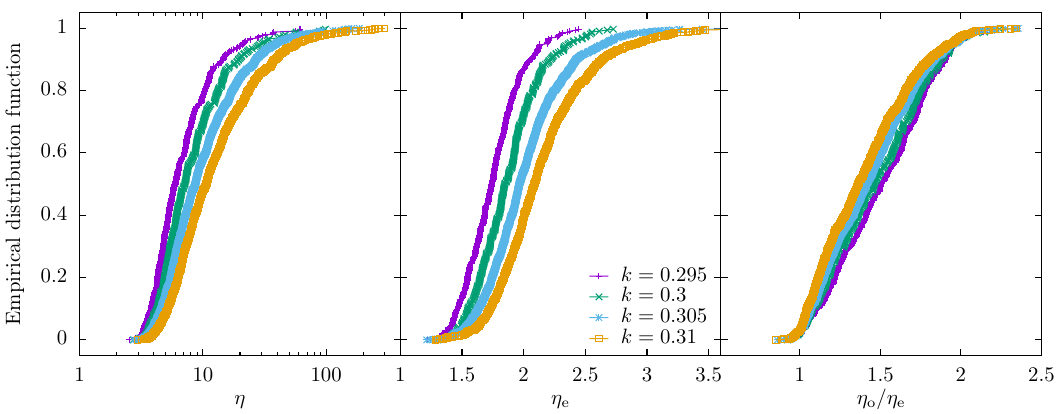}
\caption{Empirical distribution function of the different Euclidean correlation lengths presented in the system, for different values of $k$. Data from the exact diagonalization of a $L=6$ system. Data from $k=0.295$, and $k=0.3$ are calculated over $320$ samples, instead of $1280$.}
\label{fig_M:diag_distr}
\end{figure*}

The schematic representation of the spectrum in Fig.~\ref{fig_M:spectrum} is based
on the distribution functions in Figure~\ref{fig_M:diag_distr} (we typically compute the inverse distribution function, see
Sect.~\ref{sectM:fits-and-distributions} for details). 

Indeed, see Fig.~\ref{fig_M:diag_distr}-{\bf a}, the correlation length $\eta$
displays very large sample-to-sample fluctuations (to the point that a logarithmic representation is advisable) and a very strong $k$-dependence. Instead, $\eta_e$ is always a number of order one in our $L\!=\!6$ samples (Fig.~\ref{fig_M:diag_distr}-{\bf b}). Furthermore, Fig.~\ref{fig_M:diag_distr}-{\bf c}, $\eta_{\text{o}}/\eta_{\text{e}}\sim 1$
in all cases.

\begin{figure}[ht]
\includegraphics[width=\columnwidth]{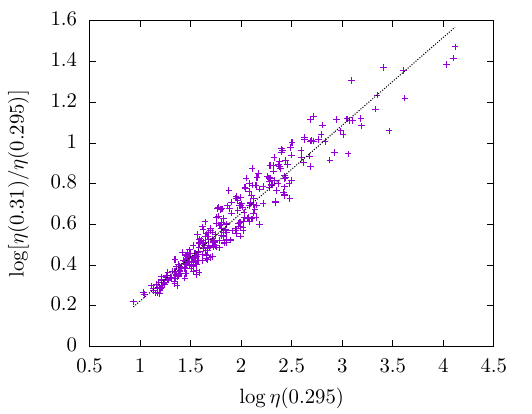}
\caption{{\bf Dependence on $k$ of the odd correlation length $\eta$.} The figure shows that the logarithm of the ratio of
$\eta(k=0.31)$ and $\eta(k=0.295)$ (computed for the same $L=6$ sample through exact diagonalization) is
very approximately a linear function of $\log \eta(k=0.295)$, with a positive slope. The figure shows data for the 350 samples that we have studied at both $k=0.295$ and $k=0.31$.}
\label{fig_M:diag_eta_k_dependence}
\end{figure}

In fact, recall Fig.~\ref{fig:odd}-{\bf c},  the distribution for $\eta$ is a Levy flight [i.e. for large $\eta$, $F(\eta)=1-\frac{B}{\eta^b}$]. The mechanism allowing exponent $b$
to vary with $k$ [hence with transverse field~\eqref{eqM:Trotter-action}] is sketched in Fig.~\ref{fig_M:diag_eta_k_dependence}. Let us compare the value of $\eta$ for the same sample at $k_1$ and $k_2$ ($k_1< k_2$). With great accuracy, $\eta(k_2)=\alpha[\eta(k_1)]^{1+\beta}$, where $\alpha,\beta$ are constants (for fixed $k_1$ and $k_2$) and $\beta>0$. This monotonic relation implies that \emph{the same} sample occupies percentile $F$ in the distribution for both $k_1$ and
$k_2$. It follows that $b(k_2)\!=\!b(k_1)/(1+\beta)$ for the exponent characterizing the Levy flight. In other words, because $b(k_2)\!<\!b(k_1)$, the tail at large $\eta$ becomes heavier as $k$ increases (see Sect.~\ref{sectM:susceptibilities} for an extended discussion).

\subsection{The critical point and critical exponents}\label{sectM:FSS}

After taking care of the $L_\tau\to\infty$ limit (within errors), in our study of the phase transition, we still need to cope with the finite spatial dimension $L$. We shall do so using Finite-Size scaling~\cite{fisher:67,fisher:72,barber:83,cardy:12}  (see Fig.~\ref{fig:GS-transition}-{\bf c}). The main questions we shall address are the computation of the critical exponents and the estimation of the critical point.  Our main tool will be the quotients method~\cite{nightingale:76,ballesteros:96,amit:05} that, surprisingly, keeps somewhat separate our two questions.

The quotients method starts by comparing a dimensionless quantity at two sizes $L_a<L_b$ (in our case, $\xi^{(3)}/L$ as a function of $k$). First, we locate a coupling $k^*{(L_a,L_b)}$ such that the curves for $L_a$ and $L_b$ cross, see Fig.~\ref{fig:GS-transition}-{\bf b}. Now, for
dimensionful quantities $A$, scaling in the thermodynamic limit as
$\xi^{x_A/\nu}$, we consider the quotient $Q_A=A_{L_a}/A_{L_b}$ at $k^*{(L_a,L_b)}$. Barring scaling corrections, $Q_A=(L_a/L_b)^{x_A/\nu}$ which yields an effective estimate of $x_A/\nu$. Indeed, considering only the leading correction to the scaling exponent, $\omega$, we have for the effective exponent
\begin{equation}\label{eqM:quotients-exponents}
\left.\frac{x_A}{\nu}\right|_{L_a,L_b}=\frac{x_A}{\nu}\,+\,\frac{1}{\log(\frac{L_b}{L_a})}\log\frac{1+D_A L_b^{-\omega}}{1+D_A L_a^{-\omega}}\,,
\end{equation}
where $D_A$ is an amplitude. Our estimates for the effective exponents can be found in Table~\ref{tabM:critical_exponents}. Yet, effective exponents need to be extrapolated to the thermodynamic limit through Eq.~\eqref{eqM:quotients-exponents}. Unfortunately, we have not been able to estimate exponent $\omega$ (there were two difficulties: first, the range of $L$ values at our disposal was small, second the analytic background~\cite{amit:05} for the $r\!=\!2$ observables and for the Binder parameter ---Sect.
\ref{sectM:one-time}--- compete with the $L^{-\omega}$ corrections). Hence, we have followed an alternative strategy. We have fitted our effective exponents to Eq.~\eqref{eqM:quotients-exponents} with fixed $\omega$ (the fit parameters were the extrapolated $x_A/\nu$ and the amplitude $D_A)$.
To account for our ignorance about $\omega$, we made it vary in a wide range $0.5\leq \omega\leq 2$. The central values in Eqs. (\ref{eq:FSS1}, \ref{eq:FSS2}) were obtained with $\omega=1$, whereas the second error estimate accounts for the $\omega$-dependence of the
$x_A/\nu$. As for the first error estimate, it is the statistical error as computed for $\omega\!=\!1$. To take into account data
we employed a bootstrap method~\cite{yllanes:11}. We considered only the diagonal part of the covariance matrix in the fits, performing a new fit for every bootstrap realization. Errors were computed from the fluctuations of the fit parameters. Fortunately, systematic errors turned out to be comparable (for $1/\nu$ smaller) with the statistical ones.

\begin{figure}[ht]
\includegraphics[width=\columnwidth]{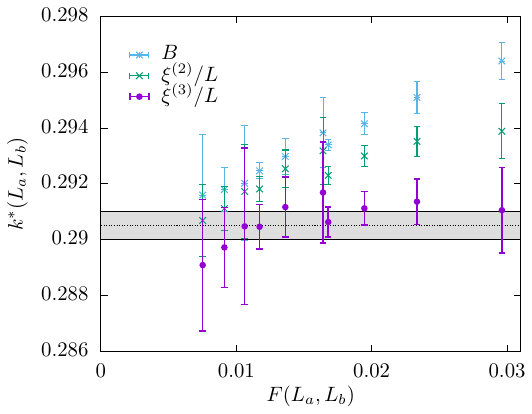}
\caption{When studied as a function of $k$ on two system sizes $L_a<L_b$, the curves for
dimensionless quantities cross at a point $k*(L_a,L_b)$, see Fig.~\ref{fig:GS-transition}-{\bf b}. The figure shows
$k^*$ (as computed for $B$, $\xi^{(2)}/L$ and $\xi^{(3)}/L$) versus
$F(L_a,L_b)$~\eqref{eqM:kc-correction}. We set $1/\nu\!=\!0.7$ and $\omega\!=\!1$ to compute $F(L_a,L_b)$. The curves should extrapolate linearly to $k_{\text{c}}$
as $F(L_a,L_b)$ tends to zero. The shaded area encompass our uncertainty in the estimation
of $k_\text{c}$.}
\label{fig_M:cortes_k_critico}
\end{figure}

As for the critical point, one expects scaling corrections of the form $k^*(L_a,L_b)=k_\text{c}+D_k F(L_a,L_b)$ ($D_k$ is an amplitude)~\cite{binder:81}:
\begin{equation}\label{eqM:kc-correction}
F(L_a,L_b)=L_a^{-(\omega+\frac{1}{\nu})}\frac{1-s^{-\omega}}{s^{1/\nu}-1}\,,\ s=\frac{L_b}{L_a}\, . 
\end{equation}
Unfortunately, this result is not of much use without a $\omega$ estimate. Fortunately, see Table~\ref{tabM:critical_exponents} and Fig.~\ref{fig_M:cortes_k_critico}, the values of $k^*(L_a,L_b)$ obtained from $\xi^{(3)}/L$ seem not to depend on size. In fact, our estimate for $k_{\text{c}}$ in Eq.~\eqref{eq:FSS1} is an interval that encompasses all our results (the shaded area in Fig.~\ref{fig_M:cortes_k_critico}). Furthermore, the crossing points for $B$ and $\xi^{(2)}/L$, see Fig.~\ref{fig_M:cortes_k_critico}, seem also reasonably well represented by Eq.~\eqref{eqM:kc-correction}.

\begin{table}[htb]
    \centering
    \renewcommand{\arraystretch}{1.5}
    \setlength{\tabcolsep}{3pt}
    \begin{tabular}{|c|c|c|c|c|c|}
    \hline
       $L_a$ & $L_b$ & $k^*$ &$1/\nu$ & $\gamma^{(2)}/\nu$ & $\gamma^{(3)}/\nu$ \\
    \hline   
$8$ & $12$ & $0.2910(15)$ & $0.47(4)$ & $1.73(7)$ & $1.07(7)$ \\
$8$ & $16$ & $0.2914(8)$ & $0.45(3)$ & $1.68(5)$ & $1.04(6)$ \\
$8$ & $20$ & $0.2911(6)$ & $0.45(2)$ & $1.68(4)$ & $0.98(3)$ \\
$8$ & $24$ & $0.2906(5)$ & $0.42(2)$ & $1.62(4)$ & $0.98(6)$ \\
$12$ & $16$ & $0.2917(18)$ & $0.43(6)$ & $1.62(11)$ & $0.99(13)$ \\
$12$ & $20$ & $0.2912(10)$ & $0.43(4)$ & $1.65(7)$ & $0.90(6)$ \\
$12$ & $24$ & $0.2905(8)$ & $0.38(3)$ & $1.55(6)$ & $0.92(9)$ \\
$16$ & $20$ & $0.290(3)$ & $0.42(12)$ & $1.7(2)$ & $0.8(2)$ \\
$16$ & $24$ & $0.2897(14)$ & $0.35(6)$ & $1.50(12)$ & $0.85(16)$ \\
$20$ & $24$ & $0.289(2)$ & $0.27(10)$ & $1.3(2)$ & $0.9(3)$ \\

    \hline
    \end{tabular}
    \caption{Crossing points $k^*(L_a,L_b)$ obtained for $\xi^{(3)}/L$ and the size-dependent, effective critical exponents [see Eq.~\eqref{eqM:quotients-exponents}] as estimated from $\partial_k \xi^{(3)}/L$ ($1/\nu$), $\chi^{(2)}$ ($\gamma^{(2)}/\nu$)  and $\chi^{(3)}$ ($\gamma^{(3)}/\nu$).  Errors are obtained using a bootstrap method.}
    \label{tabM:critical_exponents}
\end{table}

\subsection{Fitting process and Euclidean correlation length estimation}\label{sectM:fits-and-distributions}

Our aim here is to determine the relevant correlation-lengths for $C(\tau)$ and $Q_2(\tau)$ at a fixed $k$, for our $N_{\text{S}}=1280$ samples. The results will be characterized through their empirical distribution function, recall Figs.~\ref{fig:even} and~\ref{fig:odd}. Given that  $N_{\text{S}}$ is large, we need an automated approach.

The first step is estimating, for a given sample, $C(\tau)$ and $Q_2(\tau)$, as well as their standard errors, by using our 6 replicas. Now, the analysis of a noisy correlation function [such as $C(\tau)$ and $Q_2(\tau)$ ---see, e.g., Fig.~\ref{fig_M:Q2_tau}] needs a fitting window~\cite{sokal:97,janus:08b}. We chose the window upper limit as $\tau_{\text{w},f} \equiv \min_{\tau} \lbrace \tau\, |\, f(\tau) = 3.5\sigma_{f(\tau)}\rbrace$, with $f(\tau)$ either $C(\tau)$ or $Q_{2,\text{s}}(\tau) = Q_2(\tau) - Q_{2,\text{pl}}$ ($Q_{2,\text{pl}}$ is the plateau, see Fig.~\ref{fig_M:Q2_tau}), and $\sigma_{f(\tau)}$ the corresponding standard error. We need to face two problems. First, for the odd $C(\tau)$
some samples have $\tau_{\text{w},C}\geq L_\tau/2$. For these samples, $\eta\!>\! L_\tau$, and hence it is impossible to estimate, see Fig.~\ref{fig:odd}-{\bf b}. $Q_{2,\text{s}}(\tau)$ is not afflicted by this problem, see Fig.\ref{fig:even}.
Second, we need to estimate the plateau $Q_{2,\text{pl}}$. To do so, we fit  $Q_2(\tau)$ for $\tau\in[L_\tau/4,L_\tau/2]$ to a constant $Q_{2,\text{pl}}$.
In the few exceptions where this fit was not acceptable (as determined by its figure of merit $\chi^2/\text{dof}$ computed with the diagonal part of the covariance matrix), we proceeded as explained below (we used $\tau_{\text{w},Q_2}=L_\tau/2$ in those  cases).

We determined the correlation lengths through fits to
$C(\tau)=B [\text{e}^{-\tau/\eta}+\text{e}^{(\tau-L_\tau)/\eta}]$, and $Q_{2}(\tau)=Q_{2,\text{pl}} + B_{\text{e}}\text{e}^{-\tau/\eta_{\text{e}}}+B_{\text{o}}\text{e}^{-\tau/\eta_{\text{o}}}$.
The fit parameters were the amplitudes and the correlation lengths (and, for the above-mentioned exceptional samples, also $Q_{2,\text{pl}}$).  Actually, for $Q_2(\tau)$ we consider fits with  one
and with two exponential terms, keeping the fit with the smallest $\chi^2/\text{dof}$  (since we cannot tell apart which of the two correlation lengths obtained in the fit corresponds to the even gap, Fig.~\ref{fig_M:spectrum}, we shall name hereafter $\eta_{\text{e}}$ to the largest of the two). As for the lowest limit of the fitting
window, we started from $\tau_{\text{min},Q_2}=1$ and $\tau_{\text{min},C}=\tau_{\text{w},C}/10$, and kept increasing
the corresponding $\tau_{\text{min}}$ until $\chi^2/\text{dof}$ went below 0.5 for $Q_2$ [below 1 for $C(\tau)$]. 

Finally, we determine the empirical distribution function for the correlation lengths. Let $X$ be
either $\log \eta$ or $\eta_{\text{e}}$ (see below
for some subtleties regarding $\eta$). We actually compute the inverse function $X[F]$ by sorting in increasing order the $N_{\text{S}}$ values of $X$
and setting $X[F=i/N_{\text{S}}$ as the $i$-th item in the ordered list. We obtain $X[F]$ at the value of $k$ of our interest through linear interpolation of $X[F]$ computed at the two nearest value s of $k$ in the Parallel Tempering grid. To estimate errors in $X[F]$
we employ a bootstrap method with 10000 as resampling value. In each resampling, we randomly pick $N_{\text{S}}$ $X$ values (for 
the chosen sample we extract $X$ from a normal distribution
centered in $X$ as obtained from the fit and with standard deviation
the fitting error for $X$).

For $X=\log\eta$ we need to cope with the problem that we could
determine $X$ for only $N_{\text{OK}}$ of our $N_{\text{S}}$
samples. We decided to determine $X[F]$ only up to 
$F_\text{safe}\equiv (N_{\text{OK}} -4\sqrt{(N_{\text{S}}-N_{\text{OK}})N_{\text{OK}}/N_{\text{S}}})/N_{\text{S}}$ (i.e.
the maximum possible $F$ minus four standard deviates). We imposed for every bootstrap resampling that $X$ could be obtained in at least $F_{\text{safe}}N_{\text{S}}$ samples (this limitation was
irrelevant in practice).

Let us conclude by mentioning that the estimates in Eq.~\eqref{eq:u-def} were obtained through a joint fit for $\eta_{\text{e}}[F]$, with $F=0.5,0.6,0.7,0.8$ and 0.9. Errors were estimated as explained in Sect.~\ref{sectM:FSS}.

\subsection{On the magnetic susceptibilities}\label{sectM:susceptibilities}
\begin{figure*}[htb]
\includegraphics[width=\textwidth]{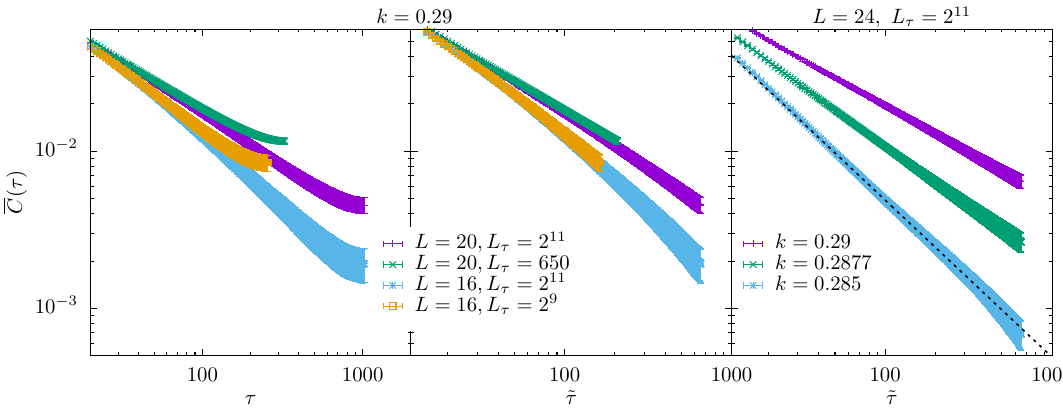}
\caption{Sample-averaged Euclidean correlation function as a function of the Euclidean distance $\tau$ (\textbf{left} panel), $\tilde{\tau}=\frac{L_{\tau}}{\pi}\sin(\pi\tau/L_{\tau})$ to avoid distortions due to the periodic boundary conditions (\textbf{center} and \textbf{right} panels). \textbf{Left} and \textbf{center} panels show the system size dependence for $k=0.29$, despite the \textbf{right} panel shows the $k$-dependence for the bigger system, $L=24$ and $L_\tau=2^{11}$. The dashed line in the \textbf{right} panel is a guide to the eye to show the critical exponent $\tilde{b}=1$ encountered for $k=0.285$ (see Sect~\ref{sectM:susceptibilities}).}
\label{fig_M:corr_eu_tau_k0285}
\end{figure*}

The sample-averaged linear susceptibility to an external magnetic field at $T=0$, $\chi^{(h)}_{\text{lin}}$, 
may diverge only if $\overline{C(\tau)}$ decays slowly for large $\tau$ [because $\chi^{(h)}_{\text{lin}}=1+2\sum_{\tau=1}^{\infty} \overline{C(\tau)}$], see Fig.~\ref{fig_M:corr_eu_tau_k0285}. Yet, the periodicity induced by the PBC, Fig.~\ref{fig_M:corr_eu_tau_k0285}-{\bf a} makes it difficult to study the behavior at large $\tau$. Fortunately, representing $\overline{C(\tau)}$ as a function of $\tilde\tau=\frac{L_\tau}{\pi}\sin (\pi\tau/L_\tau)=\tau[1+{\cal O}(\tau^2/L_\tau^2)]$ greatly alleviates this problem, see Fig.~\ref{fig_M:corr_eu_tau_k0285}-{\bf b}. Thus armed, we can study the long-time decay of $C(\tau)\propto 1/\tilde{\tau}^{\tilde{b}}$ as a function of $k$, see Fig.~\ref{fig_M:corr_eu_tau_k0285}-{\bf c}. Indeed, $\tilde{b}$ decreases as $k$ increases.
Clearly, recall that $C(\tau)\sim B\text{e}^{-\tau/\eta}$ for any sample, the mechanism discussed in
Sect.~\ref{sectM:exact-diagonalization} is at play: the heavy tail of $F(\eta)$ becomes heavier as $k$ increases, which results in a decreasing exponent $\tilde{b}$. In fact, the critical exponent $\tilde{b}=1$ is encountered at $k\!\approx\!0.285$, well into the paramagnetic phase ($\chi^{(h)}_{\text{lin}}=\infty$ if $\tilde{b}\!\leq\!1$). 

Indeed, the Levy-flight perspective provides a simple explanation for the results in Ref.~\cite{guo:94,thill:95}. In a single sample, the different susceptibilities to a magnetic field (linear, third-order, etc.) are proportional to increasing powers of $\eta$.
Hence, the existence of the disorder average of a given (generalized) susceptibility boils down to the existence of the corresponding moment of the distribution $F(\eta)$: as soon as $F(\eta)$ decays for large $\eta$ as a power-law,  some (probably a higher-order one) disorder-averaged susceptibility will diverge. Lower-order susceptibilities diverge at larger values of $k$. Hence, it is not advisable to use this approach to locate the critical point.

%

\end{document}